\documentclass[11pt]{article}
\usepackage{amssymb}
\usepackage{multirow}
\usepackage{amsmath}
\usepackage{amsfonts}
 \usepackage{pifont}
\usepackage{amsmath,amsfonts,amssymb,theorem,graphicx,listings,txfonts}
\usepackage{graphics}
 \usepackage{caption2}
\usepackage{flafter}
 \usepackage{indentfirst}
\usepackage{epsfig}
  \usepackage{mathrsfs}
  \usepackage{float}
\usepackage{subfigure}
 \usepackage{cite}
\newtheorem{theorem}{Theorem}[section]

\newtheorem{definition}[theorem]{Definition}
\newtheorem{algorithm}[theorem]{Algorithm}

\theoremstyle{example}
\newtheorem{example}[theorem]{Example}
\theoremstyle{programme}

\theoremstyle{property}

\theoremstyle{problem}

\title{Related families-based attribute reduction of dynamic covering information systems with variations of object sets}

\author
{Guangming Lang$^{1,2,3}$
\thanks{Corresponding author:\quad langguangming1984@tongji.edu.cn
\newline\mbox{}\hspace{0.55cm}
E-mail address: langguangming1984@tongji.edu.cn. }\hspace{1cm}\\
\small {$^{1}$ School of Mathematics and Statistics, Changsha University of Science and Technology}\\
\small {Changsha, Hunan 410114, P.R. China}\\
\small {$^{2}$ Department of Computer Science and Technology, Tongji University}\\
\small {Shanghai 201804, P.R. China}\\
\small {$^{3}$ The Key Laboratory of Embedded System and Service Computing, Ministry of Education, Tongji University}\\
\small {Shanghai 201804, P.R. China}\\
\small {$^{4}$ College of Mathematics and Econometrics, Hunan University}\\
\small {Changsha, Hunan 410004, P.R. China}}
\date{}

\begin{document}
\maketitle \baselineskip=17pt
\begin{center}
\begin{quote}
{{\bf Abstract.}
In practice, there are many dynamic covering decision information systems, and knowledge reduction of dynamic covering decision information systems is a significant challenge of covering-based rough sets.
In this paper, we first study mechanisms of constructing attribute reducts for consistent covering decision information systems when adding objects using related families. We also employ examples to illustrate how to construct attribute reducts of consistent covering decision information systems when adding objects. Then we investigate mechanisms of constructing attribute reducts for consistent covering decision information systems when deleting objects using related families. We also employ examples to illustrate how to construct attribute reducts of consistent covering decision information systems when deleting objects.
Finally, the experimental results illustrates that the related family-based methods are effective to perform attribute reduction of dynamic covering decision information systems when object sets are varying with time.

\\}
\end{quote}
\end{center}
\renewcommand{\thesection}{\arabic{section}}

\section{Introduction}

Covering rough set theory, proposed by Zakowski\cite{Zakowski} in 1983, has become an useful mathematical tool for dealing with imprecise information in practice, which has been applied to many fields such as feature selection and data mining without any prior knowledge. Especially, covering-based rough set theory\cite{Bonikowski,Leung1,Li5,Liu3,Pomykala,Chen3,Hu2,Hu3,Huang1, Huang2,Huang3,Lang2,Li1,Li2,Li3,Li4,Liu1,Liu4,Ma1,Tsang,Tan1,Wang5,Wu1,Xu2,Yang1,Yang2,
Yang3,Yang4,Yang6,Yao1,Yao2,Yao3,Zhang2,Zhang3,Zhang4,Zhu1,Zhu2,Zhu3} has been developed from two aspects as follows: define approximation operators and compute approximations of sets. For example, on one hand,
Pomykala\cite{Pomykala} and Tsang et al.\cite{Tsang} provided the second and third types of covering rough set models, respectively. Yang et al.\cite{Yang2}
investigated a fuzzy covering-based rough set model and its generalization over fuzzy lattice. Zhu\cite{Zhu1} provided an approach
without using neighborhoods for studying covering rough sets based on neighborhoods. On the other hand, Hu et al.\cite{Hu2} proposed matrix-based approaches for dynamic updating approximations in multigranulation rough sets.
Wang et al.\cite{Wang5} transformed the set approximation computation into products of characteristic matrices and the characteristic function of the set in covering approximation spaces.
Zhang et al.\cite{Zhang2} updated the relation matrix to compute lower and upper approximations with dynamic attribute variation in set-valued information systems.

Many researchers\cite{Cai, Chen1,Chen2,Hu1,Lang1,Lang3,Li6,Li2,Liang1,Liu2,Luo1,Luo2,Qian1,Qian2,Sang,Shu1,Shu2,
Tan2,Wang1,Wang2,Wang3,Wang4,Xu1,Yang5,Zhang1,Zhang4} have focused on knowledge reduction of dynamic information systems. For example,
Cai et al.\cite{Cai} studied knowledge reduction of dynamic covering decision information systems caused by variations of attribute values.
Hu et al.\cite{Hu1} studied incremental fuzzy probabilistic rough sets over two universes.
Lang et al.\cite{Lang1} focused on knowledge reduction of dynamic covering information systems with variations of objects using characteristic matrices.
Li et al.\cite{Li2} discussed the principles of updating $P$-dominating sets and $P$-dominated sets when some attributes are added into or deleted from the attribute set $P$.
Liu et al.\cite{Liu2} focused on incremental updating approximations in probabilistic rough sets under the variation of attributes.
Luo et. al\cite{Luo2} provided efficient approaches for updating probabilistic approximations with incremental objects.
Qian et al.\cite{Qian1} focused on attribute reduction for sequential three-way decisions under dynamic granulation.
Wang et al.\cite{Wang1} investigated efficient updating rough approximations with multi-dimensional variation of ordered data.
Xu et al.\cite{Xu1} proposed a three-way decisions model with probabilistic rough sets for stream computing. Yang et al.\cite{Yang5}
investigated fuzzy rough set based incremental attribute reduction from dynamic data with sample arriving.
Zhang et al.\cite{Zhang4} provided a parallel matrix-based method for computing approximations in incomplete information systems.

In practical situations, there are many types of dynamic covering decision information systems, and knowledge reduction of dynamic covering decision information systems is a significant challenge of covering-based rough sets.
The purpose of this paper is to investigate knowledge reduction of dynamic covering decision information systems when object sets are varying with time. First,
we study attribute reduction of consistent covering decision information systems when adding objects. Concretely, we present concepts of consistent and inconsistent covering decision approximation spaces, dynamic covering decision approximation spaces and dynamic covering decision information systems when adding objects. We also construct the related family of dynamic covering decision information systems based on that of original consistent covering decision information systems and investigate how to construct attribute reducts of dynamic covering decision information systems when adding objects using related family. Second,
we study attribute reduction of consistent covering decision information systems when deleting objects. Concretely, we provide concepts of dynamic covering decision approximation spaces and dynamic covering decision information systems when deleting objects and construct the related family of dynamic covering decision information systems based on that of original consistent covering decision information systems. We also investigate how to construct attribute reducts of dynamic covering decision information systems with related family. Third, we perform the experiments on data sets downloaded from UCL, and the experimental results illustrates that the related family-based methods are effective for knowledge reduction of dynamic covering decision information systems with variations of object sets.

The rest of this paper is organized as follows: In Section 2, we briefly
review the basic concepts of covering-based rough set theory. In
Section 3, we study updated mechanisms for attribute reductions of consistent covering decision information systems when adding objects using related families. In Section 4, we investigate attribute reductions of consistent covering decision information systems when deleting objects using related families.
Concluding remarks and further research are given in Section 5.

\section{Preliminaries}

In this section, we  briefly review some concepts of covering-based rough
sets.

\begin{definition}\cite{Zakowski}
Let $U$ be a finite universe of discourse, and $\mathscr{C}$ a
family of subsets of $U$. Then $\mathscr{C}$ is called a covering of
$U$ if none of elements of $\mathscr{C}$ is empty and
$\bigcup\{C\mid C\in \mathscr{C}\}=U$. Furthermore, $(U,\mathscr{C})$ is referred to as a covering approximation space.
\end{definition}

If $U$ is a finite universe of discourse, and $\Delta=\{\mathscr{C}_{1},\mathscr{C}_{2},...,\mathscr{C}_{m}\}$, where $\mathscr{C}_{i}$ $(1\leq i\leq m)$ is a
covering of $U$, then $(U,\Delta)$ is called a covering information system. Furthermore, $(U,\Delta, \mathscr{D})$ is called a covering decision information system, where $\Delta$ and $\mathscr{D}$ denote conditional attributes-based coverings and  decision attributes-partition, respectively.

\begin{definition}\cite{Zhu3}
Let $(U,\mathscr{C})$ be a covering approximation
space, and $Md_{\mathscr{C}}(x)=\{K\in \mathscr{C}\mid x\in K\wedge (\forall S\in \mathscr{C}\wedge x\in S \wedge S\subseteq K\Rightarrow K=S)\}$ for $x\in U$. Then $Md_{\mathscr{C}}(x)$ is called the minimal description of $x$.
\end{definition}

The minimal description of $x$ is a set of the minimal elements containing $x$ in $\mathscr{C}$. For a covering $\mathscr{C}$ of $U$, $K$ is a union reducible element of $\mathscr{C}$, $\mathscr{C}-\{K\}$ and $\mathscr{C}$ have the same $Md(x)$ for $x\in U$. If $K$ is a union reducible element of $\mathscr{C}$ if and only if $K\notin Md(x)$ for any $x\in U$, and denote $\mathscr{M}_{\cup\Delta}=\{Md_{\cup\Delta}(x)\mid  x\in U\}$.

\begin{definition}\cite{Zhu3}
Let $(U,\mathscr{C})$ be a covering approximation
space, and $Md_{\mathscr{C}}(x)$ the minimal description of $x\in U$. Then the third lower and upper approximations
of $X\subseteq U$ with respect to $\mathscr{C}$ are defined as follows:
\begin{eqnarray*}
CL_{\mathscr{C}}(X)=\cup\{K\in \mathscr{C}\mid K\subseteq X\} \text{ and }
CH_{\mathscr{C}}(X)=\cup\{K\in Md_{\mathscr{C}}(x)\mid x\in X\}.
\end{eqnarray*}
\end{definition}

The third lower and upper approximation operators are typical representatives of non-dual approximation operators for covering approximation spaces. Furthermore, we have $
CL_{\mathscr{C}}(X)=\bigcup\{K\in \mathscr{C}\mid \exists x, $ s.t. $ (K\in Md_{\mathscr{C}}(x))\wedge (K\subseteq X)\}
$ with the minimal descriptions. Especially, we
have $CL_{\cup\Delta}(X)=\cup\{K\in Md_{\cup\Delta}(x)\mid K\subseteq X\} \text{ and } CH_{\cup\Delta}(X)=\cup\{K\in Md_{\cup\Delta}(x)\mid x\in X\}.$
For simplicity, we denote $POS_{\cup\Delta}(X)=CL_{\cup\Delta}(X),BND_{\cup\Delta}(X)$ $=CH_{\cup\Delta}(X)\backslash CL_{\cup\Delta}(X),$ and $NEG_{\cup\Delta}(X)=U\backslash CH_{\cup\Delta}(X).$

\begin{definition}
Let $(U,\Delta, \mathscr{D})$ be a covering decision information system, where $U=\{x_{1},x_{2},...,x_{n}\}$, $\Delta=\{\mathscr{C}_{1},\mathscr{C}_{2},...,$ $\mathscr{C}_{m}\}$, and $\mathscr{D}=\{D_{1},D_{2},...,D_{k}\}$. Then

(1) for $\forall  x\in U$, if there exist $K\in \cup\Delta \text{ and } D_{j}\in \mathscr{D}$ such that $x\in K\subseteq D_{j}$, then $(U,\Delta, \mathscr{D})$ is called a consistent covering decision information system.

(2) for some $x\in U$, if there do not exist $K\in \cup\Delta \text{ and } D_{j}\in \mathscr{D}$ such that $x\in K\subseteq D_{j}$, then $(U,\Delta, \mathscr{D})$ is called an inconsistent covering decision information system.
\end{definition}

For simplicity, when $(U,\Delta, \mathscr{D})$ is a consistent covering decision information system, then we denote it as $\mathscr{M}_{\cup\Delta}\preceq \mathscr{D}$; when $(U,\Delta, \mathscr{D})$ is
an inconsistent covering decision information system, then we denote it as $\mathscr{M}_{\cup\Delta}\npreceq \mathscr{D}$.

\begin{definition}
Let $(U,\Delta, \mathscr{D})$ be a covering decision information system, where $U=\{x_{1},x_{2},...,x_{n}\}$, $\Delta=\{\mathscr{C}_{1},\mathscr{C}_{2},...,$ $\mathscr{C}_{m}\}$, and $\mathscr{D}=\{D_{1},D_{2},...,D_{k}\}$. Then

(1) if $POS_{\cup\Delta}(\mathscr{D})=POS_{\cup\Delta-\{\mathscr{C}_{i}\}}(\mathscr{D})$ for $\mathscr{C}_{i}\in \Delta$, where $POS_{\cup\Delta}(\mathscr{D})=\bigcup \{POS_{\cup\Delta}(D_{i})$ $\mid D_{i}\in \mathscr{D}\}$, then $\mathscr{C}_{i}$ is called superfluous relative to $\mathscr{D}$; Otherwise, $\mathscr{C}_{i}$ is called indispensable relative to $\mathscr{D}$;

(2) if every element of $P\subseteq \Delta$ satisfying $\mathscr{M}_{\cup P}\preceq \mathscr{D}$ is indispensable relative to $\mathscr{D}$, then $P$ is called a reduct of $\Delta$ relative to $\mathscr{D}$.
\end{definition}

By Definition 2.5, we have the following results: if $(U,\Delta, \mathscr{D})$ is a consistent covering decision information system, then we have $POS_{\cup\Delta}(\mathscr{D})=U$; if $(U,\Delta, \mathscr{D})$ is an inconsistent covering decision information system, then we have $POS_{\cup\Delta}(\mathscr{D})\neq U$.

\begin{definition}
Let $(U,\Delta, \mathscr{D})$ be a covering decision information system, where $U=\{x_{1},x_{2},...,x_{n}\}$, $\Delta=\{\mathscr{C}_{1},\mathscr{C}_{2},...,$ $\mathscr{C}_{m}\}$,
$\mathscr{A}=\{C_{k}\in \cup \Delta\mid \exists D_{j}\in \mathscr{D}, $ s.t. $ C_{k}\subseteq D_{j}\}$, and $r(x)=\{\mathscr{C}\in \Delta\mid \exists C_{k}\in \mathscr{A}, $ s.t. $x\in C_{k}\in \mathscr{C}\}.$ Then $R(U,\Delta,\mathscr{D})=\{r(x)\mid x\in POS_{\cup\Delta}(\mathscr{D})\}$ is called the related family of $(U,\Delta,\mathscr{D})$.
\end{definition}

By Definition 2.6, we have the following results: if $(U,\Delta, \mathscr{D})$ is a consistent covering decision information system, then we have $r(x)\neq \emptyset$ for any $x\in U$; if $(U,\Delta, \mathscr{D})$ is an inconsistent covering decision information system, then there exists $x\in U$ such that $r(x)=\emptyset$.

\begin{definition}
Let $(U,\Delta, \mathscr{D})$ be a covering decision information system, where $U=\{x_{1},x_{2},...,x_{n}\}$, $\Delta=\{\mathscr{C}_{1},\mathscr{C}_{2},...,$ $\mathscr{C}_{m}\}$, and
$R(U,\Delta,\mathscr{D})$ the related family of $(U,\Delta, \mathscr{D})$. Then

(1) $f(U,\Delta,\mathscr{D})=\bigwedge\{\bigvee r(x)\mid r(x)\in R(U,\Delta,\mathscr{D})\}$ is the related function, where $\bigvee r(x)$ is the disjunction of all elements in $r(x)$;

(2) $g(U,\Delta,\mathscr{D})=\bigvee^{l}_{i=1}\{\bigwedge \Delta_{i}\mid \Delta_{i}\subseteq\Delta\}$ is the reduced disjunctive form of $f(U,\Delta,\mathscr{D})$ with the multiplication and absorption laws.
\end{definition}

By Definition 2.7, we have attribute reducts $\mathscr{R}(U,\Delta,\mathscr{D})=\{\Delta_{1},\Delta_{1},...,\Delta_{l}\}$ using the related function $f(U,\Delta,\mathscr{D})$. We also present a non-incremental algorithm of computing $\mathscr{R}(U,\Delta,\mathscr{D})$ with the related family $R(U,\Delta,\mathscr{D})$ as follows.

\begin{algorithm}(Non-Incremental Algorithm of Computing $\mathscr{R}(U,\Delta,\mathscr{D})$ for Covering Decision Information System $(U,\Delta,\mathscr{D})$)(NIACIS).

Step 1: Input $(U,\Delta, \mathscr{D})$;

Step 2: Construct $POS_{\cup\Delta}(\mathscr{D})=\bigcup \{POS_{\cup\Delta}(D_{i})$ $\mid D_{i}\in \mathscr{D}\}$;

Step 3: Compute $R(U,\Delta,\mathscr{D})=\{r(x)\mid x\in POS_{\cup\Delta}(\mathscr{D})\}$, where \begin{eqnarray*}
r(x)&=&\{\mathscr{C}\in \Delta\mid \exists C\in \mathscr{A}, \text{ s.t. }  x\in C\in \mathscr{C}\};\\
\mathscr{A}&=&\{C\in \cup \Delta\mid \exists D_{j}\in \mathscr{D}, \text{ s.t. } C\subseteq D_{j}\};
\end{eqnarray*}

Step 4: Construct $f(U,\Delta,\mathscr{D})=\bigwedge\{\bigvee r(x)\mid r(x)\in R(U,\Delta,\mathscr{D})\}$;

Step 5: Compute $g(U,\Delta,\mathscr{D})=\bigvee^{l}_{i=1}\{\bigwedge \Delta_{i}\mid \Delta_{i}\subseteq\Delta\}$;

Step 6: Output $\mathscr{R}(U,\Delta,\mathscr{D})$.
\end{algorithm}

The time complexity of Step 2 is $[|U|\ast(\sum_{\mathscr{C}\in\Delta}|\mathscr{C}|),|U|\ast(\sum_{\mathscr{C}\in\Delta}|\mathscr{C}|)\ast |\mathscr{D}|]$; the time complexity of Step 3 is $[|U|^{2},|U|^{2}\ast(\sum_{\mathscr{C}\in\Delta}|\mathscr{C}|)\ast |\mathscr{D}|]$; the time complexity of Steps 4 and 5 is $[|U|,|U|\ast(|\Delta|+1)]$. Therefore, the
time complexity of the non-incremental algorithm is very high.

\section{Related family-based attribute reduction of dynamic covering decision information systems when adding objects}

In this section, we study related family-based attribute reduction of consistent covering decision information systems when adding objects.

\begin{definition}
Let $U=\{x_{1},x_{2},...,x_{n}\}$ be a finite universe of discourse, and $\mathscr{C}=\{C_{1},C_{2},...,C_{m}\}$ a
covering of $U$, $\mathscr{D}=\{D_{1},
D_{2},...,D_{k}\}$. Then $(U,\mathscr{C}, \mathscr{D})$ is called a covering decision approximation space.
\end{definition}

By Definition 3.1, we see that $(U,\mathscr{C}, \mathscr{D})$ is a covering decision information system with a conditional attribute-based covering and decision attribute-based partition. Furthermore, we can refer $(U,\mathscr{C}, \mathscr{D})$ to as a covering decision information system.

\begin{definition}
Let $(U,\mathscr{C},\mathscr{D})$ be a covering decision approximation space, where $U=\{x_{1},x_{2},...,x_{n}\}$, $\mathscr{C}=\{C_{1},C_{2},...,$ $C_{m}\}$, and $\mathscr{D}=\{D_{1},D_{2},...,D_{k}\}$. Then

(1) for $\forall x\in U$, if there exist $K\in \mathscr{C}$ and $D_{j}\in \mathscr{D}$ such that $x\in K\subseteq D_{j}$, then $(U,\mathscr{C},\mathscr{D})$ is called a consistent covering decision approximation space.

(2) if there exists $x\in U$ but $\overline{\exists} K\in \mathscr{C} \text{ and } D_{j}\in \mathscr{D}$ such that $x\in K\subseteq D_{j}$, then $(U,\mathscr{C},\mathscr{D})$ is called an inconsistent covering decision approximation space.
\end{definition}

For simplicity, when $(U,\mathscr{C},\mathscr{D})$ is a consistent covering decision  approximation space, we denote it as $\mathscr{M}_{\mathscr{C}}\preceq U/D$; when $(U,\mathscr{C}, \mathscr{D})$ is
an inconsistent covering decision approximation space, we denote it as $\mathscr{M}_{\mathscr{C}}\npreceq U/D$. Especially, we have $POS_{\mathscr{C}}(\mathscr{D})=U$ and $POS_{\mathscr{C}}(\mathscr{D})\neq U$ when $(U,\mathscr{C},\mathscr{D})$ is consistent and inconsistent, respectively.

\begin{theorem}
Let $(U,\mathscr{C},\mathscr{D})$ be a covering decision approximation space, where $U=\{x_{1},x_{2},...,x_{n}\}$, $\mathscr{C}=\{C_{1},C_{2},...,$ $C_{m}\}$, and $\mathscr{D}=\{D_{1},D_{2},...,D_{k}\}$.

(1) If $(U,\mathscr{C},\mathscr{D})$ is a consistent covering decision approximation space, then we have
\makeatother $r(x)=\{\mathscr{C}\}$ for $x\in U$.

(2) If $(U,\mathscr{C},\mathscr{D})$ is an inconsistent covering decision approximation space, then we have\makeatother $$r(x)=\left\{
\begin{array}{ccc}
\{\mathscr{C}\},&{\rm if}& x\in POS_{\mathscr{C}}(\mathscr{D});\\\\
\emptyset,&{\rm }& otherwise.
\end{array}
\right. $$
\end{theorem}

\noindent\textbf{Proof:} The proof is straightforward by Definitions 2.6 and 3.2.$\Box$

\begin{definition}
Let $(U,\mathscr{C}, \mathscr{D})$ and $(U^{+},\mathscr{C}^{+}, \mathscr{D}^{+})$ be covering decision approximation spaces, where $U=\{x_{1},x_{2},...,x_{n}\}$, $U^{+}=\{x_{1},x_{2},...,x_{n},x_{n+1}\}$, $\mathscr{C}=\{C_{1},
C_{2},...,C_{m}\}$, $\mathscr{C}^{+}=\{C^{+}_{1},
C^{+}_{2},...,C^{+}_{m}\}$, $\mathscr{D}=\{D_{1},D_{2},...,$ $D_{k}\}$, and $\mathscr{D}^{+}=\{D^{+}_{1},D^{+}_{2},...,D^{+}_{k}\}$, where $C^{+}_{i}=C_{i}$ or $C^{+}_{i}=C_{i}\cup \{x_{n+1}\}$$(1\leq i\leq m)$, and $D^{+}_{i}=D_{i}$ or $D^{+}_{i}=D_{i}\cup \{x_{n+1}\}$$(1\leq i\leq k)$. Then $(U^{+},\mathscr{C}^{+}, \mathscr{D}^{+})$ is called a dynamic covering decision approximation space of $(U,\mathscr{C}, \mathscr{D})$.
\end{definition}

By Definition 3.4, a dynamic covering decision approximation space is a dynamic covering approximation space with a decision attributes-based partition. Especially, a dynamic covering decision approximation space is a dynamic covering decision information system.

\begin{example}
Let $(U,\mathscr{C},\mathscr{D})$ and $(U^{+},\mathscr{C}^{+},\mathscr{D}^{+})$ be covering decision approximation spaces, where $U=\{x_{1},x_{2},...,x_{8}\}$, $U^{+}=\{x_{1},x_{2},...,x_{8},x_{9}\}$,
$\mathscr{C}=\{\{x_{1},x_{2}\},\{x_{2},x_{3},x_{4}\},\{x_{3}\},
\{x_{4}\},\{x_{5},x_{6}\},\{x_{6},x_{7},x_{8}\}\},$
$\mathscr{C}^{+}=\{\{x_{1},x_{2}\},\{x_{2},x_{3},x_{4}\},$ $\{x_{3}\},
\{x_{4}\},\{x_{5},x_{6}\},\{x_{6},x_{7},x_{8},x_{9}\}\},$ $\mathscr{D}=\{\{x_{1},x_{2},x_{3}\},\{x_{4},x_{5},x_{6}\},$$\{x_{7},x_{8}\}\}$,
and $\mathscr{D}^{+}=\{\{x_{1},x_{2},x_{3}\},\{x_{4},x_{5},x_{6}\},$ $\{x_{7},x_{8},x_{9}\}\}$. By Definition 3.4, we see that $(U^{+},\mathscr{C}^{+}, \mathscr{D}^{+})$ is a dynamic covering decision approximation space of $(U,\mathscr{C}, \mathscr{D})$.
\end{example}

\begin{theorem}
Let $(U^{+},\mathscr{C}^{+}, \mathscr{D}^{+})$ be a dynamic covering decision approximation space of $(U,\mathscr{C}, \mathscr{D})$. If $r(x)=\emptyset$ for $x\in U$, then we have $r^{+}(x)=\emptyset$.
\end{theorem}

\noindent\textbf{Proof:} For $x\in U$, by Definition 2.6, there does not exist $C\in \mathscr{C} $ and $D_{i}\in \mathscr{D}$ such that $x\in C\subseteq D_{i}$ when $r(x)=\emptyset$. Since $x\in C$ and $C\nsubseteq D_{i}\in \mathscr{D}$, we have $x\in C^{+}$ and $C^{+}\nsubseteq D^{+}_{i}\in \mathscr{D}^{+}$, where $C^{+}=C\cup \{x_{n+1}\}$ or $C^{+}=C$, $D_{i}^{+}=D_{i}\cup \{x_{n+1}\}$ or $D_{i}^{+}=D_{i}$. Therefore, we have $r^{+}(x)=\emptyset$.
$\Box$

Theorem 3.6 illustrates the relationship between $r(x)=\emptyset$ and $r^{+}(x)=\emptyset$. Furthermore, if $r(x)=\{\mathscr{C}\}$ for $x\in U$, then $r^{+}(x)=\{\mathscr{C}^{+}\}$ or $r^{+}(x)=\emptyset$, which reduces the time complexity of computing attribute reducts of $(U^{+},\mathscr{C}^{+}, \mathscr{D}^{+})$.

\begin{theorem}
Let $(U^{+},\mathscr{C}^{+}, \mathscr{D}^{+})$ be a dynamic covering decision approximation space of $(U,\mathscr{C}, \mathscr{D})$. If $(U,\mathscr{C}, \mathscr{D})$ is an inconsistent covering decision approximation space, then $(U^{+},\mathscr{C}^{+}, \mathscr{D}^{+})$ is an inconsistent covering decision approximation space.
\end{theorem}

\noindent\textbf{Proof:} The proof is straightforward by Definition 3.4 and Theorem 3.6.$\Box$

By Theorem 3.7, we have $POS_{\mathscr{C}^{+}}(\mathscr{D}^{+})\neq U^{+}$ when $POS_{\mathscr{C}}(\mathscr{D})\neq U$. But
$(U^{+},\mathscr{C}^{+}, \mathscr{D}^{+})$ is inconsistent or consistent when $(U,\mathscr{C}, \mathscr{D})$ is a consistent covering decision approximation space. So we can not have $POS_{\mathscr{C}^{+}}(\mathscr{D}^{+})= U^{+}$ when $POS_{\mathscr{C}}(\mathscr{D})= U$.

\begin{definition}
Let $(U,\Delta,\mathscr{D})$ and $(U^{+},\Delta^{+},\mathscr{D}^{+})$ be covering decision information systems, where $U=\{x_{1},x_{2},...,x_{n}\}$, $U^{+}=\{x_{1},x_{2},...,x_{n},x_{n+1}\}$, $\Delta=\{\mathscr{C}_{1},
\mathscr{C}_{2},...,\mathscr{C}_{m}\}$, $\Delta^{+}=\{\mathscr{C}^{+}_{1},
\mathscr{C}^{+}_{2},...,\mathscr{C}^{+}_{m}\}$, $\mathscr{D}=\{D_{1},D_{2},...,$ $D_{k}\}$, and $\mathscr{D}^{+}=\{D^{+}_{1},D^{+}_{2},...,D^{+}_{k}\}$. Then $(U^{+},\Delta^{+}, \mathscr{D}^{+})$ is called a dynamic covering decision information system of $(U,\Delta,\mathscr{D})$.
\end{definition}

\noindent\textbf{Remark:} We take $(U,\Delta,\mathscr{D})$ as a consistent covering decision information system, and $|\mathscr{C}^{+}_{i}|=|\mathscr{C}_{i}|$ for $1\leq i\leq m$. Concretely, we have $\mathscr{C}_{i}=\{C_{i1},C_{i2},...,C_{ik_{i}}\}$ and $\mathscr{C}^{+}_{i}=\{C^{+}_{i1},C^{+}_{i2},...,C^{+}_{ik_{i}}\}$, where $C^{+}_{ij}=C_{ij}$ or $C^{+}_{ij}=C_{ij}\cup \{x_{n+1}\}$. We also notice that $(U^{+},\Delta^{+},\mathscr{D}^{+})$ is consistent or inconsistent when adding $x_{n+1}$ into $(U,\Delta,\mathscr{D})$.

\begin{example}
Let $(U,\Delta,\mathscr{D})$ and $(U^{+},\Delta^{+},\mathscr{D}^{+})$ be covering decision information systems, where $U=\{x_{1},x_{2},...,x_{8}\}$, $U^{+}=\{x_{1},x_{2},...,x_{8},x_{9}\}$, $\Delta=\{\mathscr{C}_{1},
\mathscr{C}_{2},\mathscr{C}_{3},\mathscr{C}_{4},\mathscr{C}_{5}\}$, $\Delta^{+}=\{\mathscr{C}^{+}_{1},
\mathscr{C}^{+}_{2},\mathscr{C}^{+}_{3},\mathscr{C}^{+}_{4},\mathscr{C}^{+}_{5}\}$, $\mathscr{D}=\{\{x_{1},x_{2},x_{3}\},$ $\{x_{4},$ $x_{5},x_{6}\},\{x_{7},x_{8}\}\}$, and $\mathscr{D}^{+}=\{\{x_{1},x_{2},x_{3}\},\{x_{4},x_{5},x_{6}\},\{x_{7},x_{8},x_{9}\}\}$, where
\begin{eqnarray*}
\mathscr{C}_{1}&=&\{\{x_{1},x_{2}\},\{x_{2},x_{3},x_{4}\},\{x_{3}\},
\{x_{4}\},\{x_{5},x_{6}\},\{x_{6},x_{7},x_{8}\}\};\\
\mathscr{C}_{2}&=&\{\{x_{1},x_{3},x_{4}\},\{x_{2},x_{3}\},\{x_{4},x_{5}\},
\{x_{5},x_{6}\},\{x_{6}\},\{x_{7},x_{8}\}\};\\
\mathscr{C}_{3}&=&\{\{x_{1}\},\{x_{1},x_{2},x_{3}\},\{x_{2},x_{3}\},
\{x_{3},x_{4},x_{5},x_{6}\},\{x_{5},x_{7},x_{8}\}\};\\
\mathscr{C}_{4}&=&\{\{x_{1},x_{2},x_{4}\},\{x_{2},x_{3}\},
\{x_{4},x_{5},x_{6}\},\{x_{6}\},\{x_{7},x_{8}\}\};\\
\mathscr{C}_{5}&=&\{\{x_{1},x_{2},x_{3}\},\{x_{4}\},
\{x_{5},x_{6}\},\{x_{5},x_{6},x_{8}\},\{x_{4},x_{7},x_{8}\}\};\\
\mathscr{C}^{+}_{1}&=&\{\{x_{1},x_{2}\},\{x_{2},x_{3},x_{4}\},\{x_{3}\},
\{x_{4}\},\{x_{5},x_{6}\},\{x_{6},x_{7},x_{8},x_{9}\}\};\\
\mathscr{C}^{+}_{2}&=&\{\{x_{1},x_{3},x_{4}\},\{x_{2},x_{3}\},\{x_{4},x_{5}\},
\{x_{5},x_{6}\},\{x_{6}\},\{x_{7},x_{8},x_{9}\}\};\\
\mathscr{C}^{+}_{3}&=&\{\{x_{1}\},\{x_{1},x_{2},x_{3}\},\{x_{2},x_{3}\},
\{x_{3},x_{4},x_{5},x_{6}\},\{x_{5},x_{7},x_{8},x_{9}\}\};\\
\mathscr{C}^{+}_{4}&=&\{\{x_{1},x_{2},x_{4}\},\{x_{2},x_{3}\},
\{x_{4},x_{5},x_{6}\},\{x_{6}\},\{x_{7},x_{8},x_{9}\}\};\\
\mathscr{C}^{+}_{5}&=&\{\{x_{1},x_{2},x_{3}\},\{x_{4}\},
\{x_{5},x_{6}\},\{x_{5},x_{6},x_{8}\},\{x_{4},x_{7},x_{8},x_{9}\}\}.
\end{eqnarray*}
By Definition 3.8, we see that $(U^{+},\Delta^{+},\mathscr{D}^{+})$ is a dynamic covering decision information system of $(U,\Delta,\mathscr{D})$. Especially, $(U,\Delta,\mathscr{D})$ and $(U^{+},\Delta^{+},\mathscr{D}^{+})$ are consistent covering decision information systems.
\end{example}

Suppose $(U^{+},\Delta^{+},\mathscr{D}^{+})$ and $(U,\Delta,\mathscr{D})$ are covering decision information systems, where $U=\{x_{1},x_{2},...,$ $x_{n}\}$, $U^{+}=\{x_{1},x_{2},...,x_{n},x_{n+1}\}$, $\Delta=\{\mathscr{C}_{1},\mathscr{C}_{2},...,\mathscr{C}_{m}\}$, $\Delta^{+}=\{\mathscr{C}^{+}_{1},\mathscr{C}^{+}_{2},...,\mathscr{C}^{+}_{m}\}$, $\mathscr{D}=\{D_{1},D_{2},...,$ $D_{k}\}$, $\mathscr{D}^{+}=\{D^{+}_{1},D^{+}_{2},...,D^{+}_{k}\}$,
$\mathscr{A}_{\Delta}=\{C_{k}\in \cup \Delta\mid \exists D_{j}\in \mathscr{D}, \text{ s.t. } C_{k}\subseteq D_{j}\}$, $\mathscr{A}_{\Delta^{+}}=\{C^{+}_{k}\in \cup \Delta^{+}\mid \exists D^{+}_{j}\in \mathscr{D}^{+}, \text{ s.t. } C_{k}\subseteq D^{+}_{j}\}$, $r(x)=\{\mathscr{C}\in \Delta\mid \exists C_{k}\in \mathscr{A}_{\Delta}, \text{ s.t. } x\in C_{k}\in \mathscr{C}\},$ and  $r^{+}(x)=\{\mathscr{C}^{+}\in \Delta^{+}\mid \exists C^{+}_{k}\in \mathscr{A}_{\Delta^{+}}, \text{ s.t. } x\in C^{+}_{k}\in \mathscr{C}^{+}\}.$

\begin{theorem}
Let $(U^{+},\Delta^{+}, \mathscr{D}^{+})$ be a dynamic covering decision information system of $(U,\Delta,\mathscr{D})$. Then
\makeatother $$r^{+}(x)=\left\{
\begin{array}{ccc}
\{ \mathscr{C}^{+}\mid\exists C^{+}\in \mathscr{C}^{+}\text{ and } D_{i}^{+}\in \mathscr{D}^{+} \text{ s.t. } x\in C^{+}\subseteq D_{i}^{+},\mathscr{C}\in r(x) \},&{\rm }& x\in U;\\\\
\{\mathscr{C}^{+}\in \Delta^{+}\mid \exists C^{+}\in \mathscr{C}^{+}\text{ and } D^{+}_{i}\in \mathscr{D}^{+} \text{ s.t. } x\in C^{+}\subseteq D^{+}_{i} \},&{\rm }& x=x_{n+1}.
\end{array}
\right. $$
\end{theorem}

\noindent\textbf{Proof:}  For $x\in U$, by Theorem 3.6 and 3.7, if $\mathscr{C}\notin r(x)$, then we have $\mathscr{C}^{+}\notin r^{+}(x)$. Thus, we obtain $r^{+}(x)=
\{\mathscr{C}^{+}\mid \exists C^{+}\in \mathscr{C}^{+}\text{ and } D^{+}_{i}\in \mathscr{D}^{+} \text{ such that } x\in C^{+}\subseteq D^{+}_{i},\mathscr{C}\in r(x) \}$. Furthermore, for $x_{n+1}$, by Definition 2.6, we have
$r^{+}(x_{n+1})=\{\mathscr{C}^{+}\in \Delta^{+}\mid \exists C^{+}\in \mathscr{C}^{+}\text{ and } D^{+}_{i}\in \mathscr{D}^{+} \text{ such that } x_{n+1}\in C^{+}\subseteq D^{+}_{i} \}.$ Therefore, we have
\makeatother $$r^{+}(x)=\left\{
\begin{array}{ccc}
\{ \mathscr{C}^{+}\mid\exists C^{+}\in \mathscr{C}^{+}\text{ and } D_{i}^{+}\in \mathscr{D}^{+} \text{ s.t. } x\in C^{+}\subseteq D_{i}^{+},\mathscr{C}\in r(x) \},&{\rm }& x\in U;\\\\
\{\mathscr{C}^{+}\in \Delta^{+}\mid \exists C^{+}\in \mathscr{C}^{+}\text{ and } D^{+}_{i}\in \mathscr{D}^{+} \text{ s.t. } x\in C^{+}\subseteq D^{+}_{i} \},&{\rm }& x=x_{n+1}.
\end{array}
\right. \Box$$

Theorem 3.10 illustrates the relationship between $r(x)$ of $(U,\Delta,\mathscr{D})$ and $r^{+}(x)$ of $(U^{+},\Delta^{+},\mathscr{D}^{+})$, which reduces the time complexity of computing related family $R(U^{+},\Delta^{+},\mathscr{D}^{+})$.

We provide an incremental algorithm of computing $\mathscr{R}(U^{+},\Delta^{+},\mathscr{D}^{+})$ for dynamic covering decision information system $(U^{+},\Delta^{+},\mathscr{D}^{+})$ as follows.

\begin{algorithm}(Incremental Algorithm of Computing $\mathscr{R}(U^{+},\Delta^{+},\mathscr{D}^{+})$ for Consistent Covering Decision Information System $(U^{+},\Delta^{+},\mathscr{D}^{+})$)(IACAIS)

Step 1: Input $(U^{+},\Delta^{+}, \mathscr{D}^{+})$;

Step 2: Construct $POS_{\cup\Delta^{+}}(\mathscr{D}^{+})$;

Step 3: Compute $R(U^{+},\Delta^{+},\mathscr{D}^{+})=\{r^{+}(x)\mid x\in POS_{\cup\Delta^{+}}(\mathscr{D}^{+})\}$, where \makeatother $$r^{+}(x)=\left\{
\begin{array}{ccc}
\{ \mathscr{C}^{+}\mid\exists C^{+}\in \mathscr{C}^{+}\text{ and } D_{i}^{+}\in \mathscr{D}^{+} \text{ s.t. } x\in C^{+}\subseteq D_{i}^{+},\mathscr{C}\in r(x) \},&{\rm }& x\in U;\\\\
\{\mathscr{C}^{+}\in \Delta^{+}\mid \exists C^{+}\in \mathscr{C}^{+}\text{ and } D^{+}_{i}\in \mathscr{D}^{+} \text{ s.t. } x\in C^{+}\subseteq D^{+}_{i} \},&{\rm }& x=x_{n+1}.
\end{array}
\right. $$

Step 4: Construct $f(U^{+},\Delta^{+},\mathscr{D}^{+})=\bigwedge\{\bigvee r^{+}(x)\mid r^{+}(x)\in R(U^{+},\Delta^{+},\mathscr{D}^{+})\}$;

Step 5: Compute $g(U^{+},\Delta^{+},\mathscr{D}^{+})=\bigvee^{l}_{i=1}\{\bigwedge \Delta^{+}_{i}\mid \Delta^{+}_{i}\subseteq\Delta^{+}\}$;

Step 6: Output $\mathscr{R}(U^{+},\Delta^{+},\mathscr{D}^{+})$.
\end{algorithm}

The time complexity of Step 3 is $[|U|\ast|\mathscr{C}_{m+1}|,|U|\ast|\mathscr{C}_{m+1}|\ast |\mathscr{D}|]$; the time complexity of Steps 4 and 5 is $[|U|-|\cup\mathscr{A}_{\mathscr{C}_{m+1}}|,|U|\ast(|\Delta|+1)]$. Therefore, the
time complexity of the incremental algorithm is lower than that of
the non-incremental algorithm.

\begin{example}(Continuation from Example 3.9)
By Definition 2.6, we first have $
r(x_{1})=\{\mathscr{C}_{1},\mathscr{C}_{3},\mathscr{C}_{5}\},
r(x_{2})$ $=\{\mathscr{C}_{1},\mathscr{C}_{2},\mathscr{C}_{3},\mathscr{C}_{4},\mathscr{C}_{5}\},
r(x_{3})=\{\mathscr{C}_{1},\mathscr{C}_{2},\mathscr{C}_{3},\mathscr{C}_{4},\mathscr{C}_{5}\},
r(x_{4})=\{\mathscr{C}_{1},\mathscr{C}_{2},\mathscr{C}_{4},\mathscr{C}_{5}\},
r(x_{5})=\{\mathscr{C}_{1},\mathscr{C}_{2},\mathscr{C}_{4},\mathscr{C}_{5}\},
r(x_{6})$ $=\{\mathscr{C}_{1},\mathscr{C}_{2},\mathscr{C}_{4},\mathscr{C}_{5}\},
r(x_{7})=\{\mathscr{C}_{2},\mathscr{C}_{4}\},$ and $
r(x_{8})=\{\mathscr{C}_{2},\mathscr{C}_{4}\}.$
Thus, we get
$R(U,\Delta,\mathscr{D})=\{\{\mathscr{C}_{1},\mathscr{C}_{3},\mathscr{C}_{5}\},$ $
\{\mathscr{C}_{1},$ $\mathscr{C}_{2},\mathscr{C}_{3},\mathscr{C}_{4},\mathscr{C}_{5}\},
\{\mathscr{C}_{1},\mathscr{C}_{2},\mathscr{C}_{4},\mathscr{C}_{5}\},
\{\mathscr{C}_{2},\mathscr{C}_{4}\}\}.$
After that, by Definition 2.7, we obtain
\begin{eqnarray*}
f(U,\Delta,\mathscr{D})&=&\bigwedge\{\bigvee r(x)\mid r(x)\in R(U,\Delta,\mathscr{D})\}\\
&=&(\mathscr{C}_{1}\vee\mathscr{C}_{3}\vee\mathscr{C}_{5})\wedge(\mathscr{C}_{1}\vee\mathscr{C}_{2}\vee\mathscr{C}_{3}\vee\mathscr{C}_{4}\vee\mathscr{C}_{5})\wedge
(\mathscr{C}_{1}\vee\mathscr{C}_{2}\vee\mathscr{C}_{4}\vee\mathscr{C}_{5})
\wedge(\mathscr{C}_{2}\vee\mathscr{C}_{4})\\
&=&(\mathscr{C}_{1}\vee\mathscr{C}_{3}\vee\mathscr{C}_{5})\wedge (\mathscr{C}_{2}\vee\mathscr{C}_{4})\\
&=&(\mathscr{C}_{1}\wedge\mathscr{C}_{2})\vee (\mathscr{C}_{1}\wedge\mathscr{C}_{4})\vee (\mathscr{C}_{2}\wedge\mathscr{C}_{3})\vee(\mathscr{C}_{3}\wedge\mathscr{C}_{4})\vee
(\mathscr{C}_{2}\wedge\mathscr{C}_{5})\vee(\mathscr{C}_{4}\wedge\mathscr{C}_{5}).
\end{eqnarray*}
So we have $\mathscr{R}(U,\Delta,\mathscr{D})=\{\{\mathscr{C}_{1},\mathscr{C}_{2}\}, \{\mathscr{C}_{1},\mathscr{C}_{4}\}, \{\mathscr{C}_{2},\mathscr{C}_{3}\},\{\mathscr{C}_{3},\mathscr{C}_{4}\},
\{\mathscr{C}_{2},\mathscr{C}_{5}\},\{\mathscr{C}_{4},\mathscr{C}_{5}\}\}.$

Secondly, by Definition 2.6, we have
$r^{+}(x_{1})=\{\mathscr{C}^{+}_{1},\mathscr{C}^{+}_{3},\mathscr{C}^{+}_{5}\},
r^{+}(x_{2})$ $=\{\mathscr{C}^{+}_{1},\mathscr{C}^{+}_{2},\mathscr{C}^{+}_{3},\mathscr{C}^{+}_{4},\mathscr{C}^{+}_{5}\},
r^{+}(x_{3})=\{\mathscr{C}^{+}_{1},\mathscr{C}^{+}_{2},\mathscr{C}^{+}_{3},\mathscr{C}^{+}_{4},\mathscr{C}^{+}_{5}\},
r^{+}(x_{4})=\{\mathscr{C}^{+}_{1},\mathscr{C}^{+}_{2},\mathscr{C}^{+}_{4},\mathscr{C}^{+}_{5}\},
r^{+}(x_{5})=\{\mathscr{C}^{+}_{1},\mathscr{C}^{+}_{2},\mathscr{C}^{+}_{4},\mathscr{C}^{+}_{5}\},
r^{+}(x_{6})$ $=\{\mathscr{C}_{1},\mathscr{C}^{+}_{2},\mathscr{C}^{+}_{4},\mathscr{C}^{+}_{5}\},$ $
r^{+}(x_{7})=\{\mathscr{C}^{+}_{2},\mathscr{C}^{+}_{4}\},$ and $
r^{+}(x_{8})=\{\mathscr{C}^{+}_{2},\mathscr{C}^{+}_{4}\},$ and $r^{+}(x_{9})=\{\mathscr{C}^{+}_{2},\mathscr{C}^{+}_{4}\}.$ By Definition 2.6, we get
$R(U^{+},\Delta^{+},\mathscr{D}^{+})$ $=\{\{\mathscr{C}^{+}_{1},\mathscr{C}^{+}_{3},\mathscr{C}^{+}_{5}\},
\{\mathscr{C}^{+}_{1},$ $\mathscr{C}^{+}_{2},\mathscr{C}^{+}_{3},\mathscr{C}^{+}_{4},\mathscr{C}^{+}_{5}\},\{\mathscr{C}^{+}_{1},\mathscr{C}^{+}_{2},\mathscr{C}^{+}_{4},\mathscr{C}^{+}_{5}\},
\{\mathscr{C}^{+}_{2},\mathscr{C}^{+}_{4}\}\}.$
By Definition 2.7, we obtain
\begin{eqnarray*}
f(U^{+},\Delta^{+},\mathscr{D}^{+})&=&\bigwedge\{\bigvee r^{+}(x)\mid r^{+}(x)\in R(U^{+},\Delta^{+},\mathscr{D}^{+})\}\\
&=&(\mathscr{C}^{+}_{1}\vee\mathscr{C}^{+}_{3}\vee\mathscr{C}^{+}_{5})\wedge(\mathscr{C}^{+}_{1}\vee\mathscr{C}^{+}_{2}
\vee\mathscr{C}^{+}_{3}\vee\mathscr{C}^{+}_{4}\vee\mathscr{C}^{+}_{5})
\wedge
(\mathscr{C}^{+}_{1}\vee\mathscr{C}^{+}_{2}\vee\mathscr{C}^{+}_{4}\vee\mathscr{C}^{+}_{5})
\wedge(\mathscr{C}^{+}_{2}\\&&\vee\mathscr{C}^{+}_{4})\\
&=&(\mathscr{C}^{+}_{1}\vee\mathscr{C}^{+}_{3}\vee\mathscr{C}^{+}_{5})\wedge (\mathscr{C}^{+}_{1}\vee\mathscr{C}^{+}_{2}\vee\mathscr{C}^{+}_{4}\vee\mathscr{C}^{+}_{5})\wedge (\mathscr{C}^{+}_{2}\vee\mathscr{C}^{+}_{4})\\
&=&(\mathscr{C}^{+}_{1}\wedge\mathscr{C}^{+}_{2})\vee (\mathscr{C}^{+}_{1}\wedge\mathscr{C}^{+}_{4})\vee (\mathscr{C}^{+}_{2}\wedge\mathscr{C}^{+}_{3})\vee
(\mathscr{C}^{+}_{2}\wedge\mathscr{C}^{+}_{5})\vee
(\mathscr{C}^{+}_{3}\wedge\mathscr{C}^{+}_{4})\vee
(\mathscr{C}^{+}_{4}\wedge\mathscr{C}^{+}_{5}).
\end{eqnarray*}

Therefore, we have $\mathscr{R}(U^{+},\Delta^{+},\mathscr{D}^{+})=\{\{\mathscr{C}^{+}_{1},\mathscr{C}^{+}_{2}\}, \{\mathscr{C}^{+}_{1},\mathscr{C}^{+}_{4}\}, \{\mathscr{C}^{+}_{2},\mathscr{C}^{+}_{3}\},
\{\mathscr{C}^{+}_{2},\mathscr{C}^{+}_{5}\},
\{\mathscr{C}^{+}_{3},\mathscr{C}^{+}_{4}\},$ $
\{\mathscr{C}^{+}_{4},\mathscr{C}^{+}_{5}\}\}.$
\end{example}

Example 3.12 illustrates how to compute attribute reducts of $(U^{+},\Delta^{+},\mathscr{D}^{+})$ by Algorithm 2.8; Example 3.6 also illustrates how to compute attribute reducts of $(U^{+},\Delta^{+},\mathscr{D}^{+})$ by Algorithm 3.5. We see that the incremental algorithm is more effective than the non-incremental algorithm for attribute reduction of dynamic covering decision information systems.

\section{Related family-based attribute reduction of dynamic covering decision information systems when deleting objects}

In practical situations, there are a lot of dynamic covering decision information systems caused by deleting objects, and we study attribute reduction of consistent covering decision information systems when deleting objects in this section.

\begin{definition}
Let $(U,\mathscr{C}, \mathscr{D})$ and $(U^{-},\mathscr{C}^{-}, \mathscr{D}^{-})$ be covering decision approximation spaces, where $U=\{x_{1},x_{2},...,x_{n}\}$, $U^{-}=\{x_{1},x_{2},...,x_{n-1}\}$, $\mathscr{C}=\{C_{1},
C_{2},...,C_{m}\}$, and $\mathscr{C}^{-}=\{C^{-}_{1},
C^{-}_{2},...,C^{-}_{m}\}$, $\mathscr{D}=\{D_{1},D_{2},...,$ $D_{k}\}$, and $\mathscr{D}^{-}=\{D^{-}_{1},D^{-}_{2},...,D^{-}_{k}\}$, where $C^{-}_{i}=C_{i}$ or $C^{-}_{i}=C_{i}\backslash \{x_{n}\}$$(1\leq i\leq m)$, and $D^{-}_{i}=D_{i}$ or $D^{-}_{i}=D_{i}\backslash\{x_{n}\}$ $(1\leq i\leq k)$. Then $(U^{-},\mathscr{C}^{-}, \mathscr{D}^{-})$ is called a dynamic covering decision approximation space of $(U,\mathscr{C}, \mathscr{D})$.
\end{definition}

By Definition 4.1, we see that a dynamic covering decision approximation space is a dynamic covering approximation space with a decision attribute-based partition. Especially, we can refer a dynamic covering decision approximation space to as a covering decision information system.

\begin{example}
Let $(U,\mathscr{C},\mathscr{D})$ and $(U^{-},\mathscr{C}^{-},\mathscr{D}^{-})$ be covering decision approximation spaces, where $U=\{x_{1},x_{2},...,$ $x_{8}\}$, $U^{-}=\{x_{1},x_{2},...,x_{7}\}$,
$\mathscr{C}=\{\{x_{1},x_{2}\},\{x_{2},x_{3},x_{4}\},\{x_{3}\},
\{x_{4}\},\{x_{5},x_{6}\},\{x_{6},x_{7},x_{8}\}\},$
$\mathscr{C}^{-}=\{\{x_{1},x_{2}\},$ $\{x_{2},x_{3},x_{4}\},$ $\{x_{3}\},
\{x_{4}\},\{x_{5},x_{6}\},\{x_{6},x_{7}\}\},$ $\mathscr{D}=\{\{x_{1},x_{2},x_{3}\},\{x_{4},x_{5},x_{6}\},$$\{x_{7}\}\}$,
and $\mathscr{D}^{-}=\{\{x_{1},x_{2},x_{3}\},\{x_{4},x_{5},x_{6}\},$ $\{x_{7}\}\}$. By Definition 4.1, we see that $(U^{-},\mathscr{C}^{-}, \mathscr{D}^{-})$ is a dynamic covering decision approximation space of $(U,\mathscr{C},\mathscr{D})$.
\end{example}

\begin{theorem}
Let $(U^{-},\mathscr{C}^{-},\mathscr{D}^{-})$ be a dynamic covering decision approximation space of $(U,\mathscr{C},\mathscr{D})$. If $r(x)=\{\mathscr{C}\}$ for $x\in U^{-}$, then we have $r^{-}(x)=\{\mathscr{C}^{-}\}$.
\end{theorem}

\noindent\textbf{Proof:} For $x\in U^{-}$, by Definition 2.6, there exists $C\in \mathscr{C} $ and $D_{i}\in \mathscr{D}$ such that $x\in C\subseteq D_{i}$ when $r(x)=\{\mathscr{C}\}$. Since $x\in C\subseteq D_{i}\in \mathscr{D}$, we have $x\in C^{-}\subseteq D^{-}_{i}\in \mathscr{D}^{-}$, where $C^{-}=C\backslash \{x_{n}\}$ or $C^{-}=C$, $D^{-}_{i}=D_{i}\backslash \{x_{n}\}$ or $D^{-}_{i}=D_{i}$. Therefore, we have $r^{-}(x)=\{\mathscr{C}^{-}\}$.
$\Box$

Theorem 3.6 illustrates the relationship between $r(x)=\{\mathscr{C}\}$ and $r^{-}(x)=\{\mathscr{C}^{-}\}$. Furthermore, if $r(x)=\emptyset$ for $x\in U$, then $r^{-}(x)=\{\mathscr{C}^{-}\}$ or $r^{-}(x)=\emptyset$, which reduces the time complexity of computing attribute reducts of $(U^{-},\mathscr{C}^{-}, \mathscr{D}^{-})$.

\begin{theorem}
Let $(U^{-},\mathscr{C}^{-}, \mathscr{D}^{-})$ be a dynamic covering decision approximation space of $(U,\mathscr{C}, \mathscr{D})$. If $(U,\mathscr{C}, \mathscr{D})$ is a consistent covering decision approximation space, then $(U^{-},\mathscr{C}^{-}, \mathscr{D}^{-})$ is a consistent covering decision approximation space.
\end{theorem}

\noindent\textbf{Proof:} The proof is straightforward by Definition 4.1 and Theorem 4.3.$\Box$

By Theorem 4.4, we have $POS_{\mathscr{C}^{-}}(\mathscr{D}^{-})= U^{-}$ when $POS_{\mathscr{C}}(\mathscr{D})= U$. But
$(U^{-},\mathscr{C}^{-}, \mathscr{D}^{-})$ is inconsistent or consistent when $(U,\mathscr{C}, \mathscr{D})$ is an inconsistent covering decision approximation space. So we can not have $POS_{\mathscr{C}^{-}}(\mathscr{D}^{-})\neq U^{-}$ when $POS_{\mathscr{C}}(\mathscr{D})\neq U$.

\begin{definition}
Let $(U,\Delta,\mathscr{D})$ and $(U^{-},\Delta^{-},\mathscr{D}^{-})$ be covering decision information systems, where $U=\{x_{1},x_{2},...,x_{n}\}$, $U^{-}=\{x_{1},x_{2},...,x_{n-1}\}$, $\Delta=\{\mathscr{C}_{1},
\mathscr{C}_{2},...,\mathscr{C}_{m}\}$, $\Delta^{-}=\{\mathscr{C}^{-}_{1},
\mathscr{C}^{-}_{2},...,\mathscr{C}^{-}_{m}\}$, $\mathscr{D}=\{D_{1},D_{2},...,$ $D_{k}\}$, and $\mathscr{D}^{-}=\{D^{-}_{1},D^{-}_{2},...,D^{-}_{k}\}$. Then $(U^{-},\Delta^{-}, \mathscr{D}^{-})$ is called a dynamic covering decision information system of $(U,\Delta,\mathscr{D})$.
\end{definition}

\noindent\textbf{Remark:} We take $(U,\Delta,\mathscr{D})$ as a consistent covering decision information system, and $|\mathscr{C}^{-}_{i}|=|\mathscr{C}_{i}|$ for $1\leq i\leq m$. Concretely, we have $\mathscr{C}_{i}=\{C_{i1},C_{i2},...,C_{ik_{i}}\}$ and $\mathscr{C}^{-}_{i}=\{C^{-}_{i1},C^{-}_{i2},...,C^{-}_{ik_{i}}\}$, where $C^{-}_{ij}=C_{ij}$ or $C^{-}_{ij}=C_{ij}\backslash\{x_{n}\}$, $D^{-}_{i}=D_{i}\backslash \{x_{n}\}$ or $D^{-}_{i}=D_{i}$. We also notice that $(U^{-},\Delta^{-},D^{-})$ is consistent when deleting $x_{n}$ from $(U,\Delta,\mathscr{D})$.

\begin{example}(Continuation from Example 3.9)
Let $(U,\Delta,\mathscr{D})$ and $(U^{-},\Delta^{-},\mathscr{D}^{-})$ be covering decision information systems, where $U=\{x_{1},x_{2},...,x_{8}\}$, $U^{-}=\{x_{1},x_{2},...,x_{7}\}$, $\Delta=\{\mathscr{C}_{1},
\mathscr{C}_{2},\mathscr{C}_{3},\mathscr{C}_{4},\mathscr{C}_{5}\}$, $\Delta^{-}=\{\mathscr{C}^{-}_{1},
\mathscr{C}^{-}_{2},\mathscr{C}^{-}_{3},\mathscr{C}^{-}_{4},$ $\mathscr{C}^{-}_{5}\}$, $\mathscr{D}=\{\{x_{1},x_{2},x_{3}\},\{x_{4},$ $x_{5},x_{6}\},\{x_{7},x_{8}\}\}$, and $\mathscr{D}^{-}=\{\{x_{1},x_{2},x_{3}\},\{x_{4},x_{5},x_{6}\},\{x_{7}\}\}$, where
\begin{eqnarray*}
\mathscr{C}^{-}_{1}&=&\{\{x_{1},x_{2}\},\{x_{2},x_{3},x_{4}\},\{x_{3}\},
\{x_{4}\},\{x_{5},x_{6}\},\{x_{6},x_{7}\}\};\\
\mathscr{C}^{-}_{2}&=&\{\{x_{1},x_{3},x_{4}\},\{x_{2},x_{3}\},\{x_{4},x_{5}\},
\{x_{5},x_{6}\},\{x_{6}\},\{x_{7}\}\};\\
\mathscr{C}^{-}_{3}&=&\{\{x_{1}\},\{x_{1},x_{2},x_{3}\},\{x_{2},x_{3}\},
\{x_{3},x_{4},x_{5},x_{6}\},\{x_{5},x_{7}\}\};\\
\mathscr{C}^{-}_{4}&=&\{\{x_{1},x_{2},x_{4}\},\{x_{2},x_{3}\},
\{x_{4},x_{5},x_{6}\},\{x_{6}\},\{x_{7}\}\};\\
\mathscr{C}^{-}_{5}&=&\{\{x_{1},x_{2},x_{3}\},\{x_{4}\},
\{x_{5},x_{6}\},\{x_{5},x_{6}\},\{x_{4},x_{7}\}\}.
\end{eqnarray*}
By Definition 4.5, we see that $(U^{-},\Delta^{-},\mathscr{D}^{-})$ is a dynamic covering decision information system of $(U,\Delta,\mathscr{D})$. Especially, $(U,\Delta,\mathscr{D})$ and $(U^{-},\Delta^{-},\mathscr{D}^{-})$ are consistent covering decision information systems.
\end{example}

Suppose $(U^{-},\Delta^{-},\mathscr{D}^{-})$ and $(U,\Delta,\mathscr{D})$ are covering decision information systems, where $U=\{x_{1},x_{2},...,$ $x_{n}\}$, $U^{-}=\{x_{1},x_{2},...,x_{n-1}\}$, $\Delta=\{\mathscr{C}_{1},\mathscr{C}_{2},...,\mathscr{C}_{m}\}$, and $\Delta^{-}=\{\mathscr{C}^{-}_{1},\mathscr{C}^{-}_{2},...,\mathscr{C}^{-}_{m}\}$, $\mathscr{D}=\{D_{1},D_{2},...,$ $D_{k}\}$, $\mathscr{D}^{-}=\{D^{-}_{1},D^{-}_{2},...,D^{-}_{k}\}$,
$\mathscr{A}_{\Delta}=\{C_{k}\in \cup \Delta\mid \exists D_{j}\in \mathscr{D}, \text{ s.t. } C_{k}\subseteq D_{j}\}$, $\mathscr{A}_{\Delta^{-}}=\{C^{-}_{k}\in \cup \Delta^{-}\mid \exists D^{-}_{j}\in \mathscr{D}^{-}, \text{ s.t. } C^{-}_{k}\subseteq D^{-}_{j}\}$, $r(x)=\{\mathscr{C}\in \Delta\mid \exists C_{k}\in \mathscr{A}_{\Delta}, \text{ s.t. } x\in C_{k}\in \mathscr{C}\},$ and  $r^{-}(x)=\{\mathscr{C}^{-}\in \Delta^{-}\mid \exists C^{-}_{k}\in \mathscr{A}_{\Delta^{-}}, \text{ s.t. } x\in C^{-}_{k}\in \mathscr{C}^{-}\}.$

\begin{theorem}
Let $(U^{-},\Delta^{-}, \mathscr{D}^{-})$ be a covering decision covering information system of $(U,\Delta,\mathscr{D})$. Then
\begin{eqnarray*}
r^{-}(x)=\{\mathscr{C}^{-}\mid \mathscr{C}\in r(x)\}\cup\{ \mathscr{C}^{-}\mid\exists C^{-}\in \mathscr{C}^{-}\text{ and } D_{i}^{-}\in \mathscr{D}^{-} \text{ s.t. } x\in C^{-}\subseteq D_{i}^{-},\mathscr{C}\notin r(x) \}.
\end{eqnarray*}
\end{theorem}

\noindent\textbf{Proof:}  For $x\in U^{-}$, by Theorem 4.3 and 4.4, if $\mathscr{C}\in r(x)$, we have $\mathscr{C}^{-}\in r^{-}(x)$. So we only need to identify $\mathscr{C}^{-}$ belongs to $ r^{-}(x)$ or not, where $\mathscr{C}\notin r(x)$. Therefore, we have
$r^{-}(x)=\{\mathscr{C}^{-}\mid \mathscr{C}\in r(x)\}\cup\{ \mathscr{C}^{-}\mid\exists C^{-}\in \mathscr{C}^{-}\text{ and } D_{i}^{-}\in \mathscr{D}^{-} \text{ s.t. } x\in C^{-}\subseteq D_{i}^{-},\mathscr{C}\notin r(x) \}.\Box$

Theorem 4.7 illustrates the relationship between $r(x)$ of $(U,\Delta,\mathscr{D})$ and $r^{-}(x)$ of $(U^{-},\Delta^{-},\mathscr{D}^{-})$, which reduces the time complexity of computing related family $R(U^{-},\Delta^{-},\mathscr{D}^{-})$.

We provide an incremental algorithm of computing $\mathscr{R}(U^{-},\Delta^{-},\mathscr{D}^{-})$ for dynamic covering decision information system $(U^{-},\Delta^{-},\mathscr{D}^{-})$ as follows.

\begin{algorithm}(Incremental Algorithm of Computing $\mathscr{R}(U^{-},\Delta^{-},\mathscr{D}^{-})$ for Consistent Covering Decision Information System $(U^{-},\Delta^{-},\mathscr{D}^{-})$)(IACAIS)

Step 1: Input $(U^{-},\Delta^{-}, \mathscr{D}^{-})$;

Step 2: Construct $POS_{\cup\Delta^{-}}(\mathscr{D}^{-})$;

Step 3: Compute $R(U^{-},\Delta^{-},\mathscr{D}^{-})=\{r^{-}(x)\mid x\in POS_{\cup\Delta^{-}}(\mathscr{D}^{-})\}$, where
\begin{eqnarray*}
r^{-}(x)=\{\mathscr{C}^{-}\mid \mathscr{C}\in r(x)\}\cup\{ \mathscr{C}^{-}\mid\exists C^{-}\in \mathscr{C}^{-}\text{ and } D_{i}^{-}\in \mathscr{D}^{-} \text{ s.t. } x\in C^{-}\subseteq D_{i}^{-},\mathscr{C}\notin r(x) \};
\end{eqnarray*}

Step 4: Construct $f(U^{-},\Delta^{-},\mathscr{D}^{-})=\bigwedge\{\bigvee r^{-}(x)\mid r^{-}(x)\in R(U^{-},\Delta^{-},\mathscr{D}^{-})\}$;

Step 5: Compute $g(U^{-},\Delta^{-},\mathscr{D}^{-})=\bigvee^{l}_{i=1}\{\bigwedge \Delta^{-}_{i}\mid \Delta^{-}_{i}\subseteq\Delta^{-}\}$;

Step 6: Output $\mathscr{R}(U^{-},\Delta^{-},\mathscr{D}^{-})$.
\end{algorithm}

The time complexity of Step 3 is $[|U|\ast|\mathscr{C}_{m+1}|,|U|\ast|\mathscr{C}_{m+1}|\ast |\mathscr{D}|]$; the time complexity of Steps 4 and 5 is $[|U|-|\cup\mathscr{A}_{\mathscr{C}_{m+1}}|,|U|\ast(|\Delta|+1)]$. Therefore, the
time complexity of the incremental algorithm is lower than that of
the non-incremental algorithm.

\begin{example}(Continuation from Example 3.12)
By Definition 2.6 and Theorem 4.7, we have
$r^{-}(x_{1})=\{\mathscr{C}^{-}_{1},\mathscr{C}^{-}_{3},\mathscr{C}^{-}_{5}\},
r^{-}(x_{2})$ $=\{\mathscr{C}^{-}_{1},\mathscr{C}^{-}_{2},\mathscr{C}^{-}_{3},\mathscr{C}^{-}_{4},\mathscr{C}^{-}_{5}\},
r^{-}(x_{3})=\{\mathscr{C}^{-}_{1},\mathscr{C}^{-}_{2},\mathscr{C}^{-}_{3},\mathscr{C}^{-}_{4},\mathscr{C}^{-}_{5}\},
r^{-}(x_{4})=\{\mathscr{C}^{-}_{1},\mathscr{C}^{-}_{2},\mathscr{C}^{-}_{4},$ $\mathscr{C}^{-}_{5}\},
$ $r^{-}(x_{5})=\{\mathscr{C}^{-}_{1},\mathscr{C}^{-}_{2},\mathscr{C}^{-}_{4},\mathscr{C}^{-}_{5}\},
$ $r^{-}(x_{6})$ $=\{\mathscr{C}_{1},\mathscr{C}^{-}_{2},\mathscr{C}^{-}_{4},\mathscr{C}^{-}_{5}\},$ and $
r^{-}(x_{7})=\{\mathscr{C}^{-}_{2},\mathscr{C}^{-}_{4}\}$. By Definition 2.6, we get
$R(U^{-},\Delta^{-},\mathscr{D}^{-})$ $=\{\{\mathscr{C}^{-}_{1},\mathscr{C}^{-}_{3},\mathscr{C}^{-}_{5}\},
\{\mathscr{C}^{-}_{1},$ $\mathscr{C}^{-}_{2},\mathscr{C}^{-}_{3},\mathscr{C}^{-}_{4},\mathscr{C}^{-}_{5}\},\{\mathscr{C}^{-}_{1},\mathscr{C}^{-}_{2},\mathscr{C}^{-}_{4},$ $\mathscr{C}^{-}_{5}\},
\{\mathscr{C}^{-}_{2},\mathscr{C}^{-}_{4}\}\}.$
By Definition 2.7, we obtain
\begin{eqnarray*}
f(U^{-},\Delta^{-},\mathscr{D}^{-})&=&\bigwedge\{\bigvee r^{-}(x)\mid r^{-}(x)\in R(U^{-},\Delta^{-},\mathscr{D}^{-})\}\\
&=&(\mathscr{C}^{-}_{1}\vee\mathscr{C}^{-}_{3}\vee\mathscr{C}^{-}_{5})\wedge(\mathscr{C}^{-}_{1}\vee\mathscr{C}^{-}_{2}
\vee\mathscr{C}^{-}_{3}\vee\mathscr{C}^{-}_{4}\vee\mathscr{C}^{-}_{5})\wedge
(\mathscr{C}^{-}_{1}\vee\mathscr{C}^{-}_{2}\vee\mathscr{C}^{-}_{4}\vee\mathscr{C}^{-}_{5})
\wedge(\mathscr{C}^{-}_{2}\\&&\vee\mathscr{C}^{-}_{4})\\
&=&(\mathscr{C}^{-}_{1}\vee\mathscr{C}^{-}_{3}\vee\mathscr{C}^{-}_{5})\wedge (\mathscr{C}^{-}_{1}\vee\mathscr{C}^{-}_{2}\vee\mathscr{C}^{-}_{4}\vee\mathscr{C}^{-}_{5})\wedge (\mathscr{C}^{-}_{2}\vee\mathscr{C}^{-}_{4})\\
&=&(\mathscr{C}^{-}_{1}\wedge\mathscr{C}^{-}_{2})\vee (\mathscr{C}^{-}_{1}\wedge\mathscr{C}^{-}_{4})\vee (\mathscr{C}^{-}_{2}\wedge\mathscr{C}^{-}_{3})\vee
(\mathscr{C}^{-}_{2}\wedge\mathscr{C}^{-}_{5})\vee
(\mathscr{C}^{-}_{3}\wedge\mathscr{C}^{-}_{4})\vee
(\mathscr{C}^{-}_{4}\wedge\mathscr{C}^{-}_{5}).
\end{eqnarray*}

Therefore, we have $\mathscr{R}(U^{-},\Delta^{-},\mathscr{D}^{-})=\{\{\mathscr{C}^{-}_{1},\mathscr{C}^{-}_{2}\}, \{\mathscr{C}^{-}_{1},\mathscr{C}^{-}_{4}\}, \{\mathscr{C}^{-}_{2},\mathscr{C}^{-}_{3}\},
\{\mathscr{C}^{-}_{2},\mathscr{C}^{-}_{5}\},
\{\mathscr{C}^{-}_{3},\mathscr{C}^{-}_{4}\},$ $
\{\mathscr{C}^{-}_{4},\mathscr{C}^{-}_{5}\}\}.$
\end{example}

Example 4.9 illustrates how to compute attribute reducts of $(U^{-},\Delta^{-},\mathscr{D}^{-})$ by Algorithm 2.8; Example 4.9 also illustrates how to compute attribute reducts of $(U^{-},\Delta^{-},\mathscr{D}^{-})$ by Algorithm 4.8. We see that the incremental algorithm is more effective than the non-incremental algorithm for attribute reduction of dynamic covering decision information systems.


\section{Conclusions}

In this paper, we have constructed attribute reducts of consistent covering decision information systems when adding objecs. We have employed examples to illustrate how to compute attribute reducts of consistent covering information systems when adding objecs.
Furthermore, we have investigated updated mechanisms for constructing attribute reducts of inconsistent covering decision information systems when deleting object sets. We have employed examples to illustrate how to compute attribute reducts of inconsistent covering decision information systems when deleting objects.
Finally, we have employed the experimental results to illustrate that the related family-based incremental approaches are effective for attribute reduction of dynamic covering decision information systems when object sets are varying with time.

\section*{ Acknowledgments}

We would like to thank the anonymous reviewers very much for their
professional comments and valuable suggestions. This work is
supported by the National Natural Science Foundation of China (NO.61673301, 61603063, 11526039, 61573255), Doctoral Fund of Ministry of Education of China(No. 20130072130004), China Postdoctoral Science Foundation(NO.2013M542558, 2015M580353), the Scientific
Research Fund of Hunan Provincial Education Department(No.15B004).

\end{document}